\documentclass[12pt]{article}

\RequirePackage{epsf}
\RequirePackage{longtable}

\begin{document}

\title{A note on the cosmological constant problem}

\author{F. Darabi \\
        Department of Physics,
        Azarbaijan University of Tarbiat Moallem, Tabriz, Iran \\
        e-mail: f.darabi@azaruniv.edu }

\date{}
\maketitle

\begin{abstract}
An speculative solution for the cosmological constant problem is proposed.
It is argued that while the true quantum vacuum energy density is of the order of $M_P^4$, the observed classical vacuum energy density may be much
smaller due to the huge amount of scattering process in the vacuum state
of quantum gravity.

\end{abstract}

\begin{center}
Keywords: {Cosmological constant, vacuum energy, spacetime fluctuation.}
\end{center}

\vfill

\noindent 

\newpage

The experimental upper bound on the current value of the cosmological constant
is extremely small. On the other hand, it is usually assumed that
an {\it effective} cosmological constant describes the energy
density of the vacuum $<\rho_{vac}>$. In fact, it is believed the vacuum energy density $<\rho_{vac}>$ includes quantum field theory contributions to the effective cosmological constant
\begin{equation}\label{1}
\Lambda_{eff}=\Lambda+ \kappa <\rho_{vac}>,
\end{equation}
where $\Lambda$ is a small bare cosmological constant corresponding to the bare energy density
\begin{equation}\label{2}
\rho_{bare} \sim  10^{-10} {\rm ~erg/cm}^3\ .
\end{equation} 
The calculations show that, the quantum field theory contributions affect enormously the value of effective cosmological constant according to 
\begin{equation}\label{3}
<\rho_{vac}> \sim
M_{EW}^4  \sim 10^{47} {\rm ~erg/cm}^3\ ,
\end{equation} 
for electroweak cut off,
\begin{equation}\label{4}
<\rho_{vac}> \sim
M_{QCD}^4  \sim  10^{36} {\rm ~erg/cm}^3\ ,
\end{equation}
for QCD cut off, 
\begin{equation}\label{5}
<\rho_{vac}> \sim
M_{GUT}^4  \sim 10^{102} {\rm ~erg/cm}^3\ ,
\end{equation} 
for GUT cut off, and 
\begin{equation}\label{6}
<\rho_{vac}> \sim
M_P^4  \sim  10^{110} {\rm ~erg/cm}^3\ ,
\end{equation} 
for Planck cut off. We know the Einstein equation is a classical equation which is applied on the scales larger than the Planck scale so that one may reasonably expect the following equation to be almost valid for energy scales of electroweak, QCD, GUT, and even Planck 
\begin{equation}\label{7}
R_{\mu\nu} - {1\over 2}Rg_{\mu\nu} 
+ \Lambda_{eff} g_{\mu\nu}
= 8\pi GT_{\mu\nu}\ . 
\end{equation}
However, observational considerations requires the following Einstein equation
\begin{equation}\label{8}
R_{\mu\nu} - {1\over 2}Rg_{\mu\nu} 
+ \Lambda g_{\mu\nu}
= 8\pi GT_{\mu\nu}\, 
\end{equation}
where the small bare cosmological constant is included instead of the huge
effective one. The point is that what we observe today is in complete agreement with this latter equation, and not the former one coupled with the effective cosmological constant. But, where the quantum field theory contributions $<\rho_{vac}>$ have gone? This is the well-known cosmological constant problem \cite{Weinberg}.

Many different solutions have been proposed to answer this important
question. The relevant ones to the present paper are those using spacetime
fluctuations. Based on a common belief, the {\it spacetime} at very short distances, namely beyond the classical domain, has enormous fluctuations. The basic idea is that gravity is an interaction with many of the same fundamental properties as the other fundamental interactions in Nature. This means that the state of this field is, at some level, uncertain and described by quantum mechanics, so the spacetime itself is also subject to the kinds of uncertainty required by quantum systems. This indeterminacy means that one cannot know with infinite precision both the geometry of spacetime, and the rate of change of the spacetime geometry, in direct analogy with Heisenberg's Uncertainty Principle for quantum systems. According to Wheeler, this indeterminacy for spacetime at the so-called Planck Scale of $10^{-33}$ centimeters and $10^{-43}$ seconds requires that spacetime has a foam like structure with sudden changes in its geometry and topology into a wealth of complex shapes and textures such as quantum black holes, quantum wormholes and even quantum baby universes \cite{Wheeler}. Hawking was the first who speculated that quantum fluctuations in spacetime
topology at small scales may play an important role in shifting the cosmological
constant to zero \cite{Hawking}. Since then, other people have been following this line of thought to solve the cosmological constant problem using quantum
wormholes as the candidate for spasetime fluctuations which can lead the cosmological constant to vanish \cite{Other}. 

In the following, we shall try to find an alternative (speculative) solution for this problem. Suppose the true vacuum energy corresponding to all non-gravitational
interactions down to the Planck cut off is the same as $<\rho_{vac}>$ outlined above. Therefore, similar to every other energy sources, this cosmological vacuum energy (or cosmological constant) has to be quantized. In ordinary quantum field theory, namely in the flat spacetime, such a huge energy can be easily observed because the flat spacetime is {\it transparent} for the non-gravitational vacuum energy fluctuations. Actually, in ordinary quantum field theory, this vacuum energy is regarded as an apparent divergence which is usually avoided by renormalization techniques. However, in the presence of gravity which physically feels every kind and component of energy through
Einstein equation (\ref{7}), we can not simply put away this huge energy by using the renormalization techniques, rather we should find a solution to the confusing problem that why we do not see such a huge energy in front of us, within Einstein equation (\ref{8}). 

To this end, we note that apart from quantum field theory there exist {\it quantum gravity} and a foamy spacetime as described above can be considered as a candidate for a quantum gravitational vacuum \cite{Other}. As in every quantum field theory where the vacuum energy is just considered as an arbitrary level of energy and can be easily discarded by normal ordering, we may consider the energy content of this foamy spacetime as an arbitrary level of energy with respect to the excitation of quantum gravity and so set it to zero, with
no loss of generality. Put it another way, suppose there would be no interaction in Nature but the quantum gravitational one. So, all the energy content of its vacuum, namely the spacetime
foam, had to be set to zero. Otherwise, we never could define and establish a classical spacetime as excitations of this vacuum state. Therefore,
the cosmological constant problem arises just due to the {\it non-zero} non-gravitational
quantum field theory contributions to the {\it zero-energy} quantum gravitational vacuum, namely one has to be worry just about the appearance of non-zero non-gravitational contributions in the quantum gravitational vacuum with
zero energy.

In this regard, one may simply think that the vacuum energy corresponding
to the non-gravitational interactions is quantized like the black body radiation. However, unlike the flat spacetime discussed above, every quantum of this
vacuum energy is now engaged with an enormous spacetime fluctuations and extensively scattered by them. This is similar to the situation of a photon, falling inside a cavity through a hole, with no chance of emerging again. It is then reasonable that the large number of spacetime fluctuations in each ${\rm cm}^3$ plays the role of a cavity for the quanta of non-gravitational vacuum energy. The probability for a single quantum of this vacuum energy to escape from being scattered in this {\it quantum spacetime foam} toward the classical spacetime is very small and so a very tiny part of the vacuum energy density $<\rho_{vac}>$ has the chance to escape from this {\it opaque} quantum foam toward the classical domain of spacetime where Einstein equation holds. This phenomena is also very similar to the situation of the {\it last scattering surface} in cosmology before which every photon inside has a very small probability to escape outward due to the continuous scattering of photons by the thermal fluctuations of elementary particles, so that all the radiation is trapped in a hot opaque region preventing an outside observer to see it. In the similar way, the opaque quantum spacetime foam screens the energy density of the non-gravitational vacuum fluctuations $<\rho_{vac}>$ through the huge and continuous scattering, and this energy density becomes almost hidden for a classical observer looking at Einstein equation (\ref{8}). Only a few quanta of non-gravitational vacuum energy that are able to scape of the quantum opaque foam (quantum gravity domain) toward the classical domain (Einstein gravity) may produce the energy density $\sim 10^{-10}~ {\rm ~erg/cm}^3$ in agreement with observations. In simple speculative words, like a flat glass which is transparent for the visible light, the flat spacetime is transparent for the non-gravitational vacuum energy; and like a rough glass which is opaque for the visible light, the spacetime foam is opaque for the non-gravitational vacuum energy. In the first case, we deliberately hide the non-gravitational vacuum energy by renormalization techniques, while in the second case the quantum gravitional vacuum itself, as spacetime fluctuations, hides the non-gravitational
vacuum energy from being observed by a classical observer. The huge non-gravitational vacuum energy is trapped inside the quantum gravitational vacuum, namely spacetime foam, and can not penetrate to the outside classical spacetime
but a tiny part. Therefore, the cosmological constant problem may be solved in this context. 

We may also try to find out if the cosmological constant has variation with
the expansion of the universe. According to the above discussion, it turns out that for a given density of non-gravitational vacuum fluctuations the more density of spacetime fluctuation, the more scattering process, and hence the more suppression of the cosmological constant. We know that Einstein equation is cosmologically described by co-moving observers. So, for a co-moving observer the density of physical spacetime fluctuations increases by the physical expansion of the universe while the density of non-gravitational vacuum fluctuations remains intact. Therefore, for this observer who considers the Einstein equation starting from Big Bang to the current epoch, the rate of penetration of non-gravitational vacuum energy trapped in spacetime foam toward classical spacetime is highly suppressed due to the increase in the co-moving density of spacetime fluctuations and accordingly the cosmological constant decreases by the expansion of the universe.

\end{document}